\documentclass[traditabstract]{aa} 

\usepackage{graphicx}
\usepackage{txfonts}
\usepackage{natbib}
\bibpunct{(}{)}{;}{a}{}{,} 
\begin{document}
\title{The location of the Crab pulsar emission region: Restrictions on synchrotron emission models}


 \author{C.-I. Bj\"ornsson, A. Sandberg \& J. Sollerman} 

\institute{The Oskar Klein Centre, Department of Astronomy, Stockholm University, AlbaNova, 106 91 Stockholm, Sweden.\\
	\email{bjornsson@astro.su.se}
             }

\titlerunning{The Crab pulsar emission region}
   \date{}
   
   \abstract{
Recent observations of the Crab pulsar show no evidence for a spectral break in the infrared regime. It is argued that the observations are consistent with a power-law spectrum in the whole observable infrared - optical range. This is taken as the starting point for an evaluation of how self-consistent incoherent synchrotron models fare in a comparison with observations. Inclusion of synchrotron self-absorption proves important as does the restriction on the observed size of the emission region imposed by the relativistic beaming thought to define the pulse profile. It is shown that the observations can be used to derive two independent constraints on the distance from the neutron star to the emission region; in addition to a direct lower limit, an indirect measure is obtained from an upper limit to the magnetic field strength. Both of these limits indicate that the emission region is located at a distance considerably larger than the light cylinder radius. The implications of this result are discussed and it is emphasized that, in order for standard incoherent synchrotron models to fit inside the light cylinder, rather special physical conditions need to be invoked.
}


\maketitle

\section{Introduction}
Incoherent synchrotron radiation was recognized early on as a likely emission mechanism for the infrared-optical pulses in the Crab pulsar \citep[e.g.,][]{S70, P71}. However, as discussed by \citet{OS70} and \citet{EP73}, some of the model constraints imposed by the pulsed nature of the emission were initially not explicitly included; e.g., the assumption that the pulses are due to a combination of rotation and relativistic streaming/small pitch angles implies that the frequency of the observed emission should be at least as large as the Doppler boosted cyclotron frequency. This puts an upper limit to the magnetic field in the emission region. With a dipolar magnetic field structure, this results in a minimum distance to the emission region corresponding roughly to the light cylinder radius or even somewhat larger \citep[e.g.,][]{golden00b}.

The spectral characteristics in the infrared-optical range have long been a matter of debate. The discussion has centered on two issues \citep[see, e.g.,][ hereafter SS09]{ss09}; namely, the value of the spectral index $\alpha_\nu$, defined such that the flux $F(\nu) \propto \nu^{\alpha_\nu}$, in the optical and the existence of a possible break and/or bump in the infrared. The main question regarding $\alpha_\nu$ is whether or not it is consistent with a value of $1/3$. Since this value is the largest possible for optically thin incoherent synchrotron radiation, such a spectrum would indicate a distribution of electron energies with a sharp low energy cut-off and that the typical synchrotron frequencies for these electrons lie above the optical frequency band. In addition to its implication for the acceleration process, it has consequences for the deduced upper limit of the magnetic field in the emission region. The value discussed above assumes that the observed frequencies correspond to the typical synchrotron frequencies, i.e., that the observed spectral range is determined by the range of electron energies. Hence, the lower frequency limit corresponds to electrons being non-relativistic in the frame where there is no streaming (i.e., the average pitch-angle is $\pi/2$). This is in contrast to the case of a low energy cut-off for which the lowest energy electrons can be highly relativistic also in the no-streaming frame. Since the inertial mass of the electron increases with the Lorentz factor, the synchrotron spectrum extends below the cyclotron frequency. Hence, the arguments leading to the upper limit of the magnetic field and the associated lower limit of the distance to the emission region are no longer valid and, {\it a priori}, no restrictions regarding the location of the emission region can be set.

The synchrotron self-absorption frequency is an important parameter for restricting the properties of the emission region. This was used already by \citet[][ see also \citealt{PS83}]{S70}  together with the expected brightness temperature to estimate the source size, i.e., its lateral extent. Simple scaling relations have also been derived for the expected emission outside the infrared-optical range for the Crab pulsar itself \citep[e.g.,][]{PS87} as well as for the infrared-optical emission in other pulsars. In the latter case, the synchrotron self-absorption frequency plays a central role \citep{oconnor05}.

Another important constraint in the synchrotron scenario is the relation, imposed by the geometry of the emission region, between the observed lateral extent of the source and its distance from the neutron star. Although some aspects of this have been considered, for example by \citet{PS87} in their scaling relations, a more extensive discussion including synchrotron self-absorption still seems to be lacking. Motivated by recent observational progress regarding the spectral properties of the Crab pulsar, it is the aim of the present paper to provide such an analysis. 

In Sect.~\ref{Obs} the observational situation is summarized and evaluated. It is concluded that $\alpha_\nu = 1/3$ is consistent with, although not required by, the observations. Furthermore, it is emphasized that recent observations
show no indications of either a break or a bump in the infrared. The temporal structure of the Crab pulses is also briefly discussed. Standard synchrotron theory is applied to a few different settings for the emission region in Sect.~\ref{Synch}. It is shown that the observations can be used to derive two independent constraints on the properties of the emission region; the first is an upper limit to the strength of the magnetic field, even in the case of a low energy cut-off in the electron energy distribution, while the second one is a direct lower limit of the distance to the emission region. Although the various settings give somewhat different limits, they all suggest that the distance to the emission region is considerably larger than the light cylinder radius. The implications of these results are discussed in Sect.~\ref{Disc}. It is concluded that an incoherent synchrotron radiation scenario is still tenable although severely restricted. A few alternative origins for the infrared-optical emission are also suggested. Throughout this paper, $cgs$-units are used.

\section{Observations}\label{Obs}

\subsection{Spectral energy distribution}\label{SED}

In this section we review some of the observations of the Crab pulsar 
spectral energy distribution (SED) in the infrared-optical regime.
We start by re-stating the lack of evidence for a spectral 
break in the observed infrared. 
As mentioned by SS09, since the early observations by
\citet[][]{middleditch83}, the existence of such a break has permeated
pulsar textbooks and literature. This has had large consequences for the
theoretical interpretation of the synchrotron emission, as seen  e.g., in
\citet[][]{oconnor05} and \citet[][]{crusius01}.

\begin{figure}
\centering
\resizebox{\hsize}{!}{\includegraphics{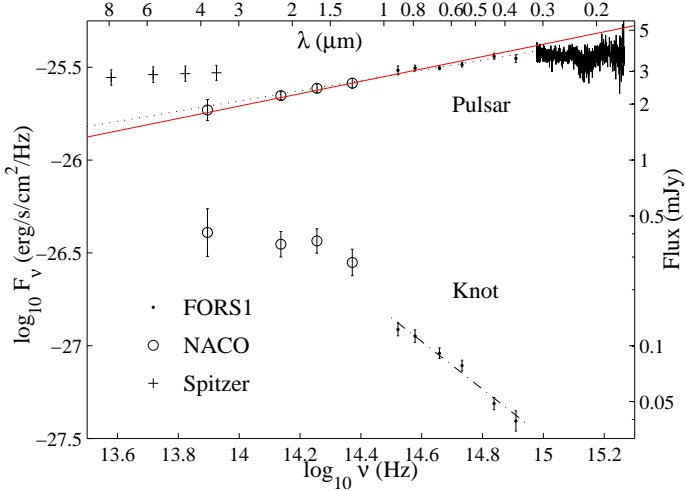}}
\caption{
Best fit of an $\alpha_\nu=1/3$ spectral index to the knot-subtracted optical and near-IR points of SS09. The lower spectrum is for the nearby knot, located 0$\farcs$6 from the pulsar. Points mark $UBVRIz$ FORS1 data, circles $JHK_sL'$ NACO data and crosses data from the {\it Spitzer Space Telescope}. The spectrum on the right hand side is the HST UV spectrum from S00. Figure adopted from SS09.      
}
\label{fig:ss09}
\end{figure}

More recent observations of the infrared SED discussed 
a 'smooth levelling' or a 'rollover' towards the IR \citep{temim06,sollerman03}.
However, looking afresh at the new observations (SS09), in particular the
corrected data from the {\it Spitzer Space Telescope} 
\citep{temim09}, there is actually no 
evidence for
any spectral break in the observed infrared (Fig.~\ref{fig:ss09}). 

For characterising the SED, we list in
Table~\ref{tab:previous_obs} results from some of 
the previous observations of the
spectral index, $\alpha_\nu$, of the Crab pulsar, and indicate 
for which wavelength ranges these observations were made 
\cite[see also][]{fordham02}.  
A quick glance at
Table~\ref{tab:previous_obs} shows that there are many observed values
of $\alpha_\nu$, on both sides of a totally "flat" spectrum. 
There are a few more observations which are not included in Table 1, 
e.g. those by  \cite{middleditch83}, 
since these data were not well fit by a power-law.

The observations behind the estimated spectral indices in 
Table~\ref{tab:previous_obs} have been obtained using many different
techniques, and this can explain at least part of the differences.
The
procedure for dereddening also has strong effects on the deduced values.

\citet[][ hereafter S00]{sollerman00} derived a value 
for the reddening towards the Crab pulsar
of
$E(B-V)=0.52$~mag, by smoothing out the 2200 $\AA$ dust feature in the near-UV,
and showed that within reasonable limits their spectral index 
had error bars of $\pm0.2$ due to the uncertainty in extinction.

\begin{figure}
\centering
\resizebox{\hsize}{!}{\includegraphics{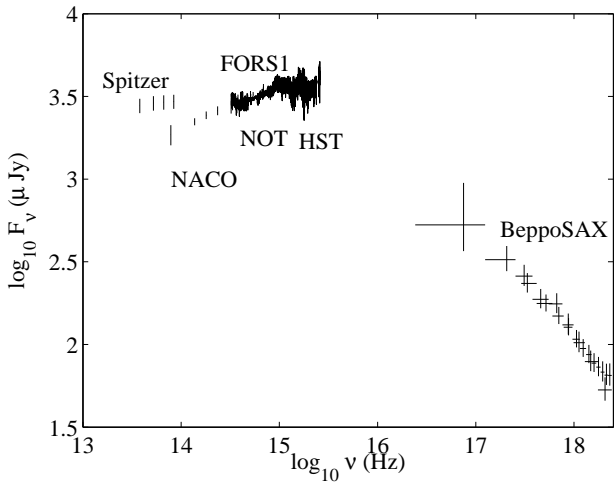}}
\caption{Multi-wavelength spectrum of the 
Crab pulsar. 
This figure is 
partly adopted from \citet{serafimovich04}, 
with the data from SS09 added.}
\label{fig:multicrab}
\end{figure}

A general trend among the measurements in Table~\ref{tab:previous_obs}
is a lower value of $\alpha_\nu$ for higher frequencies. 
S00 derived $\alpha_\nu \approx 0.11$ 
for the UV-optical range whereas the UV alone
had a spectral index that was essentially flat.
The spectral energy distribution of the Crab pulsar thus seems to level out
(and peak, see Fig.~\ref{fig:multicrab}) somewhere in the ultraviolet, and the wavelength range will
therefore also be an important factor for the derived value of
$\alpha_\nu$. 

Another large uncertainty in determining the shape of the pulsar spectrum
comes from the difficulties in 
subtracting the nebular contribution. Some of the measurements have
used time-resolved data to estimate the phase-averaged spectral index, 
while most observations have been time-integrated and therefore more 
susceptible to bad seeing.
In e.g., \citet[][]{beskin01}, the spectrum
display clear residuals from nebular subtraction.
Since the surrounding nebula appears redder than the pulsar at these wavelengths
\citep[e.g.][]{tziamtzis09}, observations at mediocre seeing may
result in a redder spectral index.
In particular, 
all time-integrated estimates 
for $\alpha_\nu$ also include contributions from the nearby
($\sim0\farcs6$) knot \citep{hester95}. 
The knot is known to be a very "red" feature ($\alpha_\nu \sim -1$), and also to be variable 
over time
\citep[][ SS09]{sollerman03}. 
The knot may thus also have affected these
measurements in different ways, depending on how much of the knot has
been included, and how strong this feature was at the time of
observation. 

It is not clear if the knot could contribute significantly to the
emission seen in the mid-IR regime.
By extrapolating a conservative spectral index for the knot
of $\alpha_\nu=-1.0$ into this regime, we can estimate that the pulsar optical
spectral index continues at least to a frequency of $\mathrm{log}~
\nu \sim 13.6$, and we will use this number below as an upper limit for the 
synchrotron self-absorption frequency ($\nu_{\rm amax}$). 

We note that there is an offset in Fig.~\ref{fig:ss09}
between the infrared flux measured with 
ground-based observations using NACO (SS09) and the space-based 
{\it Spitzer} data. 
This is likely due to 
problems with absolute flux calibration between different instruments.
The complex background inside the Crab Nebula make the resolution
important for correctly sampling the contribution of the spatially
varying 
nebular background close to
the pulsar.  Clearly, the actual shape of the SED from the
{\it Spitzer} measurements is less affected than the  
absolute flux.

We argue, thus, that the simplest possible picture for the 
knot-subtracted pulsar SED is that it can
be described as a single power-law
below the UV (optical towards infrared), and that the value of
$\alpha_\nu \sim 0.3$, is consistent with the
multitude of published observations of the Crab pulsar.
The best value in the
optical-IR regime is $\alpha_\nu=0.27\pm0.03$ (SS09), where
emission from the knot has been subtracted. 
Figure~\ref{fig:ss09} shows the best fit of a spectral index of
$\alpha_\nu=1/3$ to the observations of 
SS09,
including the
near-UV spectrum from S00.

\begin{table*}
\centering
\begin{tabular}{c c c}
\hline\hline
Reference & $\alpha_\nu$ & $\lambda$-range ($\AA$) \\
\hline 
\cite{nasuti96}	& $-0.10\pm0.01$	& $4900-7000$  \\ 
\cite{carraminana00} & $0.2\pm0.1$		& $5000-7500$  \\
\cite{sollerman00}	& $0.11\pm0.04$		& $1140-9250$  \\ 
\cite{beskin01}		& $-0.15\pm0.15$	& $5000-7050$   \\ 
\cite{fordham02}	& $-0.034\pm0.009$ & $4600-6000$  \\
\hline
\cite{percival93} 	& $0.11\pm0.13$ 	& $1680-7400$  \\ 
\cite{golden00b}	& $-0.07\pm0.19$	& $3700-5500$   \\ 
\cite{sollerman03} & $0.31\pm0.02$ & $12400-21800$  \\
\cite{ss09} 		& $0.27\pm0.03$ & $3700-38000$ \\
\hline
\end{tabular}
\caption{Summary of previous observations of the spectral index $\alpha_\nu$ of the Crab pulsar. 
The uppermost part are spectroscopic observations and the lower part are photometric. For dereddening E(B$-$V)=0.51-0.52 and $R = 3.1$ were used for all observations. 
Compare \citet{fordham02}, their Table 6. 
 }
\label{tab:previous_obs}  
\end{table*}

Figure~\ref{fig:multicrab} shows a multi-wavelength phase-averaged 
spectrum of the Crab pulsar. From this spectrum, we have estimated the
peak of the emission to lie at a frequency of 
$\mathrm{log}~\nu_{p}\approx15.6$. 
This peak frequency is also used below to derive 
constraints on the emission model.

\subsection{Temporal resolution of the pulse}

Another restriction on pulsar emission theory that is 
important for our purposes 
is given
by the temporal extent and shape of the pulse peaks.
Early data from \cite{papaliolios70} described a cusp-like behaviour of the 
pulse at a resolution of 32 $\mu$s. More recent data, with a much better time resolution of only 
$\sim1~\mu$s, instead give an approximation
of the temporal extent of the plateau on the main pulse of $\sim$ 55
$\mu$s  \citep{golden00b}.

\section{Incoherent synchrotron models}\label{Synch}
The spectral shape of the emission from the Crab pulsar is thus consistent with being due to incoherent synchrotron radiation from a distribution of relativistic electrons with a sharp low energy cut-off. In order for the rotation of the neutron star to give rise to pulses, relativistic beaming is necessary. We denote the frame, in which the total momentum of the relativistic electrons is zero (i.e., the no-streaming frame), with a prime ($'$) and let $\Gamma$ be the Lorentz factor of this frame as measured in the observers frame. 

A few standard approximations are useful to simplify the treatment. (1) The electron distribution is isotropic and mono-energetic in the primed frame with an electron energy corresponding to a Lorentz factor $\gamma_{\rm me}'$. (2) The intensity leaving the source is confined within a cone with opening angle $1/\Gamma$. Inside this cone the intensity is constant. (3) Frequencies in the two frames are related by $\nu = \Gamma \nu '$. Furthermore, $\Gamma \gg 1$ will be assumed.

 The observed intensity is then \citep[e.g.,][]{P70}
 \begin{equation}
	I(\nu) = \frac{3}{2} m \nu^{2} \gamma [1-\exp(-\tau)],
	\label{eq:1.1}
\end{equation}
where $\gamma \equiv \Gamma \gamma_{\rm me}'$ is the Lorentz factor of the radiating electrons as measured in the observers frame, $\tau$ is the synchrotron optical depth and $m$ is the mass of the electron. It will be assumed that the emission is produced at a distance $R$ from the neutron star. Radiation from the streaming electrons is observed from a surface of size  $a \pi R^2 / \Gamma^2$. The value of $a$ is unity for radially streaming electrons and a spherically symmetric source. In many cases, the value of $a$ is likely to be smaller than unity, either due to an actual source size smaller than $R/\Gamma$ or the source geometry. The observed flux can then be written
\begin{eqnarray}
	F(\nu) &=& \frac{a \pi R^2}{\Gamma^2 d^2} I(\nu) \nonumber\\
	            &=& \frac{3a \pi R^2}{2 d^2} m \nu^2 \frac{\gamma_{\rm me}'}{\Gamma}[1-\exp(-\tau)],
	\label{eq:1.2}
\end{eqnarray}
where $d$ is the distance to the pulsar.

\subsection{Emission from within the light cylinder}\label{Synch1}
Since the magnetic field is dynamically dominant in this case, the particles are streaming along the magnetic field lines with a Lorentz factor $\Gamma$. Let $R_{\rm c}$ denote the radius of curvature of the magnetic field. The distance over which particles and radiation can interact is then $2 R_{\rm c}/\Gamma$, which implies a column density of particles given by $2R_{\rm c} n / \Gamma$, where $n$ is the number density of particles. However, due to the relativistic streaming of the particles, the emitted photons experience only a fraction $1/2\Gamma^2$ of this column density. The optical depth, which is a relativistic invariant, is \citep[e.g.,][]{P70}
\begin{equation}
	\tau = \frac{4 G(5/3)}{3^{5/6} (2 \pi)^{2/3} \sin \theta} \frac{e^{8/3} B^{2/3}}
	{\left(mc \gamma_{\rm me}' \nu' \right)^{5/3}} \frac{n'R}{\Gamma^2}.
	\label{eq:1.3}
\end{equation}
Here, $R_{\rm c} = 4R/3 \sin \theta$ has been used, where $\theta$ is the angle from the magnetic axis to the emission region. Furthermore, $e$ is the electron charge and $G (z)$ is the factorial/gamma function. 

With the use of $\nu = \Gamma \nu'$, the optical depth can also be expressed as
\begin{equation}
	\tau = b^{5/3} \frac{\epsilon_{\rm e}'}{\epsilon_{\rm B}} \frac{\nu_{\rm B}^{8/3}}{\nu^{5/3}}
	\frac{R}{c (\gamma_{\rm me}')^{8/3} \Gamma^{1/3}},
	\label{eq:1.4}
\end{equation}
where $b \equiv (2\pi G(5/3)/3^{5/6} \sin \theta)^{3/5}$. Here, $\epsilon _{\rm e}'$ is the energy density of electrons, $\epsilon_{\rm B}$ is the energy density of the magnetic field, which is an invariant for Lorentz boosts along the magnetic field lines, and $\nu_{\rm B} \equiv eB/2\pi mc$ is the cyclotron frequency. Defining the self-absorption frequency as $\tau (\nu_{\rm abs}) \equiv 1$, leads to
\begin{equation}
	\nu_{\rm abs} = b \left(\frac{\epsilon _{\rm e}}{\epsilon_{\rm B}}\right)^{3/5} 
	\frac{\nu_{\rm B}^{8/5}}{\Gamma^{7/5} (\gamma_{\rm me}')^{8/5}} 
	\left(\frac{R}{c}\right)^{3/5},
	\label{eq:1.5}
\end{equation}
where $\epsilon_{\rm e} = \Gamma^2 \epsilon_{\rm e}'$ is the energy density of electrons as measured in the observers frame.
Introducing $\nu_{\rm m} \equiv \nu_{\rm B} \Gamma (\gamma_{\rm me}')^2$, equation (\ref{eq:1.5}) can be written
\begin{equation}
	\nu_{\rm abs} = b \left(\frac{\epsilon _{\rm e}}{\epsilon_{\rm B}}\right)^{3/5} 
	\frac{\nu_{\rm B}^{12/5}}{\Gamma^{3/5} \nu_{\rm m}^{4/5}} 
	\left(\frac{R}{c}\right)^{3/5}.
	\label{eq:1.6}
\end{equation}

The value of $\nu_{\rm m}$ is related to the frequency where the observed flux peaks ($\nu_{\rm p}$) through $\nu_{\rm p} \approx 0.25 \nu_{\rm m}$. When the electron distribution is not mono-energetic but rather a power-law with a low energy cut-off at $\gamma_{\rm me}'$, the value of $\nu_{\rm p}$ depends on the power-law index but is typically a few times larger than for the mono-energetic case.

The flux at $\nu_{\rm abs}$ is obtained from
\begin{equation}
	4\pi d^2 \nu_{\rm abs} F(\nu_{\rm abs}) = L_{\rm o}\left(\frac{\nu_{\rm abs}}{\nu_{\rm m}}
	\right)^{4/3},
	\label{eq:1.7}
\end{equation}
where $L_{\rm o}$ is the fiducial isotropic peak luminosity of the pulsar. With the use of equation (\ref{eq:1.2}), this can be written
\begin{equation}
	L_{\rm o} = 3.8 \pi^2 a m R^2 \frac{\nu_{\rm m}^{11/6} \nu_{\rm abs}^{5/3}}
	 {\Gamma^{3/2} \nu_{\rm B}^{1/2}}.
	 \label{eq:1.8}
\end{equation}

Expressions for $R$ and $\nu_{\rm B}$ are obtained from equations (\ref{eq:1.6}) and (\ref{eq:1.8}) as
\begin{equation}
	R = \frac{c}{\left(3.8 \pi^2 a mc^2\right)^{8/17}b^{5/51}} 
	\frac{L_{\rm o}^{8/17} \Gamma^{13/17}}{\nu_{\rm m}^{40/51} \nu_{\rm abs}^{35/51}}
	\left(\frac{\epsilon_{\rm B}}{\epsilon_{\rm e}}\right)^{1/17}
	\label{eq:1.9}
\end{equation}
and
\begin{equation}
	\nu_{\rm B} = \frac{\left(3.8 \pi^2 a mc^2\right)^{2/17}}{b^{20/51}}
	\frac{\nu_{\rm m}^{9/17} \nu_{\rm abs}^{10/17}\Gamma^{1/17}}{L_{\rm o}^{2/17}}
	\left(\frac{\epsilon_{\rm B}}{\epsilon_{\rm e}}\right)^{4/17}.
	\label{eq:1.10}	
\end{equation}
For the values of $\nu_{\rm abs}$ and $\Gamma$ we have only upper and lower limits, respectively, i.e., $\nu_{\rm abs} < \nu_{\rm amax}$ and $\Gamma > \Gamma_{\rm min}$. The lower limit to $\Gamma$ accounts for the fact that the observed pulse profile may be influenced by effects in addition to relativistic streaming.
It is convenient to normalize the value of $R$ to that of the light cylinder radius ($R_{\rm LC} = 1.6\times 10^8$) and the value of $\nu_{\rm B}$ to its value at the light cylinder (i.e., $\nu_{\rm B,LC} = 2.8\times 10^{12}$, where $B=10^6$ at the light cylinder has been assumed). With the use of $R \equiv R_{\rm norm} R_{\rm LC}$ and $\nu_{\rm B} \equiv \nu_{\rm B,norm} \nu_{\rm B,LC}$, equations (\ref{eq:1.9}) and (\ref{eq:1.10}) can be written
\begin{equation}
	R_{\rm norm} > \frac{4.4}{a^{8/17}}\frac{L_{\rm o,36}^{8/17} 
	\Gamma_{\rm min}^{12/17}} {\nu_{\rm m,15}^{40/51} \nu_{\rm amax,13}^{35/51}}
	\left(\frac{\epsilon_{\rm B} \Gamma \sin \theta}{\epsilon_{\rm e}}\right)^{1/17}
	\label{eq:1.11}
\end{equation}
and
\begin{equation}
	\nu_{\rm B,norm} < 2.0 \times 10^{-2}  a^{2/17}
	\frac{\nu_{\rm m,15}^{9/17} \nu_{\rm amax,13}^{10/17}}{L_{\rm o,
	36}^{2/17}\Gamma_{\rm min}^{3/17}}
	\left(\frac{\epsilon_{\rm B} \Gamma \sin \theta}{\epsilon_{\rm e}}\right)^{4/17}.
	\label{eq:1.12}	
\end{equation}
Here, $\nu_{\rm m} \equiv 10^{15} \nu_{\rm m,15}$, $\nu_{\rm abs} \equiv 10^{13} \nu_{\rm amax,13}$ and $L_{\rm o} \equiv 10^{36} L_{\rm o,36}$.

As discussed above, $a \lesssim 1$ is expected. From the discussion in Sect.~\ref{Obs}, $\nu_{\rm amax,13} \approx 4.0$ and the observed minimum pulse structure of $\sim$ 55
$\mu$s corresponds to $\Gamma_{\rm min} \approx 10^2$ \citep{golden00b}. It is likely that the X-ray flux is a continuation of the optical emission (see Fig. ~\ref{fig:multicrab}), i.e., that $\gamma_{\rm me}'$ is the low energy cut-off of a power-law distribution of electron energies. Hence, $\nu_{\rm m} \approx \nu_{\rm p}$ and Fig.~\ref{fig:multicrab} suggests $\nu_{\rm m,15} \approx 4.0$ for a $\nu^{1/3}$-spectrum in the optical frequency range. The peak flux in a single pulse was estimated from Fig.~\ref{fig:multicrab} together with the pulse profile in S00 to be $43$ mJy at $\nu = 10^{15}$ Hz. Extrapolating this value to $\nu_{\rm m}$ results in $L_{\rm o,36} \approx 1.4$. When these values are inserted in equation (\ref{eq:1.11}), one obtains
\begin{equation}
	R_{\rm norm} \gtrsim 1.7 \times 10\left(\frac{\epsilon_{\rm B} \Gamma \sin \theta}{\epsilon_{\rm e}}
	\right)^{1/17}.
	\label{eq:1.13}
\end{equation}
The value of $\epsilon_{\rm B} \sin \theta / \epsilon_{\rm e}$ is not known. Even though small values of $\theta$ cannot be excluded, the small exponent in equation (\ref{eq:1.13}) together with the expectation that within the light cylinder $\epsilon_{\rm B} > \epsilon_{\rm e}$ should make $R_{\rm norm} > 10$ a robust conclusion. Although the value of $\nu_{\rm m}$ is obtained by interpolating between the optical and X-ray regimes, the limit on $R_{\rm norm}$ ($\propto L_{\rm o}^{8/17}/\nu_{\rm m}^{40/51} \propto \nu_{\rm m}^{-8/51}$) is not too sensitive to its actual value for a $\nu^{1/3}$-spectrum. Likewise,
\begin{equation}
	\nu_{\rm B,norm} \lesssim 4.0\times 10^{-2}\left(\frac{\epsilon_{\rm B} \Gamma \sin \theta}
	{\epsilon_{\rm e}}\right)^{4/17}.
	\label{eq:1.14}
\end{equation}
The conclusion regarding the upper limit for the value of $\nu_{\rm B,norm}$ is not as robust as the lower limit for $R_{\rm norm}$, since the corresponding exponent is larger than that in equation (\ref{eq:1.13}) and $\epsilon_{\rm B} \gg \epsilon_{\rm e}$ cannot be excluded. However, it should be noted that increasing the upper limit of $\nu_{\rm B,norm}$ by invoking large values of $\epsilon_{\rm B} \Gamma \sin \theta / \epsilon_{\rm e}$ also increases the distance to the emission region. With the use of equations (\ref{eq:1.13}) and (\ref{eq:1.14}), this restriction can be expressed as
\begin{equation}
	R_{\rm norm} \gtrsim 3.7 \times 10 \,\nu_{\rm B,norm}^{1/4}.
	\label{eq:1.14a}
\end{equation}
Hence, independent of the detailed physics (i.e., the value of $\epsilon_{\rm B} \Gamma \sin \theta / \epsilon_{\rm e}$), it is not possible to self-consistently move the emission site to a distance corresponding to the light cylinder.

One implicit assumption underlying the results in equations (\ref{eq:1.13}) and (\ref{eq:1.14}) is that the pulse profile is independent of frequency. Although the minimum pulse structure as well as the Full Width Half Maximum (FWHM) are consistent with being constant in the UBV-range, there is an indication of a decreasing FWHM with frequency \cite[e.g.,][]{golden00b}. The suggested variation is small enough to be caused by the relativistic streaming itself, since it induces an anti-correlation between frequency and angular width. Hence, if real, this variation could indicate that the value of $\Gamma$ is not much larger than $\Gamma_{\rm min}$. Furthermore, a pulse profile varying with frequency would result in an observed spectrum flatter than the intrinsic one due to overlapping emission regions. Since observations of the Crab pulsar are consistent with a $\nu^{1/3}$-spectrum, this limits the magnitude of possible frequency variations of the pulsar profile in a synchrotron scenario. 

It should also be remembered that the low frequency part of the pulsar spectrum shown in Fig.~\ref{fig:ss09}, and, hence, the value of $\nu_{\rm amax}$ used above, is from {\it Spitzer}-observations. Since these data cannot resolve individual pulses, it could be that this emission is not pulsed. Although there are no indications that this should be the case, a conservative approach is therefore to use only the NACO-data, since \cite{eikenberry97} have shown that the pulse profiles in $JHK$ are similar to the optical ones. Taking the K-band observations as the lower limit to the pulsed emission would increase the value of $\nu_{\rm amax}$ by a factor $\approx 3$. As can be seen from equations (\ref{eq:1.11}) and (\ref{eq:1.12}), the limits of $R_{\rm norm}$ and $\nu_{\rm B,norm}$ are then changed only by a factor $\approx 2$. Hence, neither the possible frequency dependence of the pulsar profile nor the unconstrained nature of the emission measured by {\it Spitzer} is likely to substantially change the conclusion about the location of the emission region. 

The above discussion assumed that $\gamma_{\rm me}'$ is either the Lorentz factor of a mono-energetic electron distribution or the low energy cut-off in a power-law distribution of electron energies. The reason for this choice is that the observed infrared-optical flux from the Crab pulsar increases with frequency consistent with a $\nu^{1/3}$-spectrum. However, as discussed in Sect.~\ref{Obs} a flatter spectrum cannot be excluded. The brightness temperature at $\nu_{\rm abs}$ is proportional to $\gamma_{\rm me}'$. A spectrum flatter than $\nu^{1/3}$ would, therefore, decrease the brightness temperature at $\nu_{\rm abs}$, since it will be determined by electrons with Lorentz factors $\gamma < \gamma_{\rm me}'$. Furthermore, for a given value of peak luminosity (i.e., $L_{\rm o}$), the flux below the peak frequency (i.e., $\nu_{\rm m}$) would be larger than in the $\nu^{1/3}$ case. Both of these effects result in an increase of the lower limit of $R_{\rm norm}$ as well as a decrease of the upper limit of $\nu_{\rm norm}$. Since a $\nu^{1/3}$-spectrum is the hardest possible for optically thin synchrotron radiation, a spectrum flatter than this would strengthen the above conclusions.

\subsection{Emission from the light cylinder}\label{Synch.3}
An alternative site for the pulsed emission is close to the light cylinder. The physical setting is here considerably more uncertain than for emission from within the light cylinder; for example, the energy densities in particles and electric field are expected to be of the same magnitude as that in the magnetic field. This could affect the emission process as well as the beaming of the radiation in several ways. Both of these effects are crucially dependent on the details of the particle motion. In order to allow a simplified discussion of these issues, the motion of the particles will be divided into three components: (1) Random motion in the zero-momentum (i.e., primed) frame in addition to the gyration of the particles around the magnetic field ($u_{\rm rand}$). (2) Streaming or bulk motion as measured in the frame co-rotating with the neutron star ($u_{\rm co}$). In the observer's frame, this component differs from that in Sect.~\ref{Synch1} mainly due to the importance of aberration effects, which make it likely that the motion is at large angles to the magnetic field. (3) The motion due to the rotation of the neutron star ($u_{\rm rot}$). 

The origin of $u_{\rm rand}$ could be small scale irregularities in the magnetic and/or the electric field. Such deviations from pure gyration in the large scale magnetic field affect mainly the low frequency emission from an individual particle. An example of such a situation is "jitter" radiation \citep{med00}, which corresponds to small scale magnetic irregularities. As shown by \citet{med00}, the low frequency part of the spectrum is determined by the statistical properties of the magnetic field; in particular, the spectral distribution is expected to differ substantially from the synchrotron case (e.g., for "jitter" radiation $\alpha_{\nu} \approx 1$ as compared to $\alpha_{\nu} = 1/3$ for the synchrotron case). Although the statistical properties of possible small scale irregularities close to the light cylinder are hard to predict, it is unlikely that they would result in a spectral index close to that of synchrotron radiation. Hence, the observation of a spectral index for the Crab pulsar close to $\alpha_{\nu} = 1/3$ suggests that $u_{\rm rand}$ does not seriously affect the emission from individual particles and, in particular, that the basic frequency is the cyclotron one. 

A situation where $u_{\rm co}$ plays a significant role is similar to the one discussed in Sect.~\ref{Synch1}. The main differences are: (1) The angle $\theta$ does not correspond to the angular distance of the emission site from the magnetic axis; instead, $\sin \theta$ is now a free parameter determining the value of $R_{\rm c}$. (2) The energy density associated with the magnetic field ($\epsilon_{\rm B}$) is not necessarily an invariant even for Lorentz boosts along the magnetic field lines. Both of these effects are contained in the factor $\epsilon_{\rm B} \Gamma \sin \theta / \epsilon_{\rm e}$ (cf. eqns [\ref{eq:1.11}] and [\ref{eq:1.12}]). As already emphasized, by suitably varying the value of this factor, the limit for either $R_{\rm norm}$ or $\nu_{\rm B,norm}$ can be made to come closer to that expected for a source distance corresponding to the light cylinder radius; however, this occurs at the expense of the other limit, which will then correspond to an even larger distance to the emission region (cf. eq. [\ref{eq:1.14a}]). Hence, the physical differences in this case are not expected to affect the main conclusions from  Sect.~\ref{Synch1}.

When the effects of $u_{\rm co}$ are negligible, the pulsar profile is determined mainly by rotation and $\Gamma = (1-R^2/R^2_{\rm LC})^{-1/2}$.
Since the emission site is now moving towards the observer, the observed duration of a pulse is
\begin{equation}
	t_{\rm obs} \approx \frac{R_{\rm LC}}{c \Gamma^3} + \frac{\Delta R}{c},
	\label{eq:1.17}
\end{equation}
where $\Delta R$ is the extension of the source in the azimuthal (i.e., rotational) direction. For $\Delta R < R_{\rm LC}/\Gamma^3$, this leads to $\Gamma \approx (R_{\rm LC}/c t_{\rm obs})^{1/3}$. Likewise, $\Delta R > R_{\rm LC}/\Gamma^3$ implies $t_{\rm obs} \approx \Delta R/c$ or $\Gamma > (R_{\rm LC}/c t_{\rm obs})^{1/3}$; hence, $\Gamma_{\rm min} \approx (R_{\rm LC}/c t_{\rm obs})^{1/3}$. This value is smaller than in the streaming scenario, where $\Gamma_{\rm min} \approx R_{\rm LC}/c t_{\rm obs}$. In outer gap models, the emission site is usually in the vicinity of the last closed magnetic field lines. This suggests that the extension of the emitting surface along the light cylinder could be as large as $R_{\rm LC}$. Rotation implies that the value of $\Gamma$ varies significantly over distances larger than $R_{\rm LC}/\Gamma^2$ in the perpendicular (i.e., radial) direction. Hence, the limits on the emitting surface is the same as in the streaming case, i.e., $a \lesssim 1$ is expected. A minor change, as compared to the streaming scenario, is also that for $\Delta R < R_{\rm LC}/\Gamma^3$ the opacity is smaller by a factor $\Delta R \Gamma^3/ R_{\rm LC}$. This effect is also absorbed in the value of $\epsilon_{\rm B} \Gamma \sin \theta / \epsilon_{\rm e}$ (cf. eq. [\ref{eq:1.3}]).

For an emission site in the vicinity of the last closed magnetic field lines, one expects $R_{\rm norm} \approx 1$ and $\nu_{\rm B,norm} \approx 1$. 
In the general case when both $u_{\rm co}$ and $u_{\rm rot}$ contribute to the effective value of $\Gamma$, the result is expected to fall in between the two extremes discussed above. It is clear that the lower value of $\Gamma_{\rm min}$ in the rotational scenario contributes most to diminish the discrepancies  between the expected values for $R_{\rm norm}$ and $\nu_{\rm B,norm}$ and their lower and upper limits, respectively, derived from observations; for example, the smaller value of $\Gamma_{\rm min}$ reduces the lower limit of $R_{\rm norm}$ by a factor $\approx 10$ and decreases only marginally the upper limit on $\nu_{\rm B,norm}$. However, even in this case, the lower limit of $R$ is a factor $\approx 2$ larger than $R_{\rm LC}$, while the upper limit of the magnetic field is more than a factor $10$ smaller than expected for an emission site close to the light cylinder.  Hence, even with the low value of $\Gamma_{\rm min}$ allowed by the rotational scenario, it is not straightforward to make an emission site close to the light cylinder consistent with observations.

\section{Discussion} \label{Disc}
Several explicit scenarios for the emission site of the Crab pulsar have been developed 
in the past, including the outer gap model
\citep{cheng86} and the polar cap model \cite[e.g.,][]{harding81}. In this paper we have instead focused on the fundamental constraints inherent to any synchrotron emission model.
The main result from Sect.~\ref{Synch} is that in the incoherent synchrotron scenario, the implied distance to the emission region is considerably larger than the light cylinder radius. This conclusion is supported by two independent pieces of evidence; namely, a direct lower limit of the distance and an indirect one through the upper limit to the magnetic field strength. Most of the uncertain physics is contained in one parameter, which affects these limits in the same way. Hence, by changing its value, one of the limits will indicate a smaller distance to the emission site, while, at the same time, the other limit would suggest an even larger distance. It is therefore not possible to make both limits compatible with a distance comparable to or smaller than the light cylinder radius by invoking a particular value for this parameter (cf. eq. [\ref{eq:1.14a}]). Furthermore, these limits are quite robust; in particular, both of them are valid using just one of two independent observations: (1) The upper limit on the self-absorption frequency ($\nu_{\rm amax}$) or (2) the upper limit on the spectral index (i.e., a $\nu^{1/3}$-spectrum).

Although the distance to the emission region may be much larger than the light cylinder radius (i.e., $R_{\rm norm} \gg 1$), it can still be situated within the light cylinder. This requires a small inclination angle ($i$) between the rotation and magnetic axes. The allowed values for $i$ are bounded from below ($i > 1/ \Gamma_{\rm min}$) to assure pulsed emission and from above ($i < 1/R_{\rm norm}$) by the requirement that the emission region lies within the light cylinder. These inequalities are satisfied in the streaming scenario, since $\Gamma_{\rm min} > R_{\rm norm}$ is needed in order for the lateral extent of the emission region to be smaller than the light cylinder radius. It is seen from equation (\ref{eq:1.11}) that $\Gamma_{\rm min} > R_{\rm norm}$ is possible for $\Gamma_{\rm min} \approx 10^2$. Hence, a streaming scenario together with a small inclination angle are compatible with observations but certainly for inclinations much smaller than typically envisioned in most models for an oblique rotator \cite[e.g.,][]{cheng86}. 

The constraints imposed by observations on an emission region located close to the light cylinder are qualitatively similar to those corresponding to streaming along the magnetic field lines within the pulsar magnetosphere. In models with the emission site in the vicinity of the last closed magnetic field lines, $R_{\rm norm} \approx 1$ is expected; hence, small inclination angles cannot be invoked to make such scenarios compatible with observations. The limits discussed above for the distance to the emission region hinge on the size of the emitting surface and it was argued that an upper bound to this size could be found in the streaming scenario. A similar upper bound is likely to apply also to emission sites close to the light cylinder, in which case such models would be untenable. However, the magnetospheric properties in the vicinity of the light cylinder remain rather uncertain, which leaves open the possibility that this upper bound could be exceeded.

The restrictions on potential incoherent synchrotron models are quite severe. The main problem afflicting them is that their maximum brightness temperature is too low to easily fit the emission region inside the light cylinder. The brightness temperature can be increased by invoking coherent /amplified radiation. An alternative that would preserve many of the attractive features of the standard synchrotron scenario is, therefore, coherent/amplified synchrotron radiation. The conditions needed for synchrotron radiation to be amplified within a pulsar magnetosphere have been discussed  by \citet{Sto82}. In this case, the emission would likely come from well within the light cylinder and a connection to the radio emission is possible. Another possibility is emission regions located outside the light cylinder.  \citet{ler70} has considered a scenario, in which the dipole field of the rotating neutron star induces oscillations at the interface with an external plasma. Although the properties of the synchrotron-like emission expected from such models have not been worked out in any detail, their applicability should not be constrained by the size of the emission region.

\begin{acknowledgements}
This research was supported by grants from the Swedish Natural 
Science Research Council.
The Oskar Klein Centre is also funded by the Swedish Natural 
Science Research Council.
The Dark Cosmology Centre is funded by the Danish National Research Foundation. 
JS is a Royal Swedish Academy of Sciences Research Fellow supported by a 
grant from the Knut and Alice Wallenberg Foundation.
We would also like to thank the referee, Marco Salvati, for a thorough reading of the manuscript as well as many constructive comments, which helped to improve the paper substantially.

\end{acknowledgements}

\end{document}